\documentclass[superscriptaddress,aps,article,frontmatterverbose,onecolumn,amsmath,amssymb]{revtex4}

\usepackage{graphicx}
\usepackage{color}
\usepackage[colorlinks,bookmarks=false,citecolor=blue,linkcolor=blue,urlcolor=blue]{hyperref}
\usepackage{ulem}

\begin{document}


%
%

\title{Ballistic Anisotropic Magnetoresistance \\ in Core Shell Nanowires and Rolled-up Nanotubes}

\author{Ching-Hao Chang}
\email{cutygo@gmail.com}
\affiliation{Institute for Theoretical Solid State Physics, IFW Dresden, Helmholtzstrasse 20 Dresden, 01069, Germany}

\author{Carmine Ortix}
\email{c.ortix@ifw-dresden.de}
\affiliation{Institute for Theoretical Solid State Physics, IFW Dresden, Helmholtzstrasse 20 Dresden, 01069, Germany}
\affiliation{Institute for Theoretical Physics, Center for Extreme Matter and Emergent Phenomena, Utrecht University, Princetonplein 5, 3584 CC, Utrecht, Netherlands}

\date{\today}

\begin{abstract}
In ferromagnetic nanostructures, the ballistic anisotropic magnetoresistance (BAMR)  is a change in the ballistic conductance with the direction of magnetization due to spin-orbit interaction. Very recently, a directional dependent ballistic conductance has been predicted to occur in a number of newly synthesized nonmagnetic semiconducting nanostructures subject to externally applied magnetic fields, without necessitating spin-orbit coupling. In this article, we review past works on the prediction of this BAMR effect in core-shell nanowires and rolled-up nanotubes. This is complemented by new results we establish for the transport properties of tubular nanosystems subject to external magnetic fields. 
\end{abstract}

\maketitle
\newpage
\section{Introduction}
The anisotropic magnetoresistance (AMR) is a generic magnetotransport property of ferromagnetic metals first discovered by Thomson\cite{tho57} back in 1857. In general, the longitudinal resistance of a bulk polycrystalline ferromagnetic metal shows maximum resistivity when the current is parallel to the magnetization direction, and minimum resistivity when the current is perpendicular to the magnetization direction. The resistance indeed only depends on the relative orientation of the magnetization vector ${\bf m}$ and the current density ${\bf j}$, and can be cast in the general form\cite{mcg75}
\begin{equation*}
\rho=\rho_{0} + \delta \rho \left( {\hat{\bf j}} \cdot {\hat{\bf m}} \right)^2, 
\end{equation*}
where $\rho_0$ is the bulk isotropic resistivity whereas $\delta \rho$ quantifies the amplitude of the AMR effect. 
The subsequent discovery four decades ago of a few percent AMR in transition metals and their alloys has led to the development of AMR sensors for magnetic recording\cite{oha00,haj08}. 
The physical origin of the AMR can be individuated in the anisotropy of scattering produced by spin-orbit interaction\cite{smi51}. A stronger scattering is indeed expected for electrons travelling parallel to the magnetization, which results in a larger resistivity $\rho_0 + \delta \rho$. 

It was subsequently proposed that this effect might also arise in ferromagnetic quasi-one-dimensional nanostructures with characteristic structural dimensions less than the electronic mean free path\cite{vel05}. In these systems, the electronic transport is ballistic rather than diffusive, and therefore electronic scattering does not contribute to the conductance. The ballistic conductance of a magnetic nanowire with width of the order of the Fermi wavelength $\lambda_F$ is indeed simply given by $G= N e^2/h$ where $N$ is the total number of open conducting channels\cite{lan88}. Because of the spin-orbit interaction, which is known to be enhanced in open and constrained geometries, the number of transverse modes at the Fermi energy changes with the magnetization direction, thereby yielding a ballistic anisotropic magnetoresistance (BAMR). The first experimental evidence for the occurrence of the BAMR effect has been provided by observing a stepwise variation in the ballistic conductance of cobalt nanocontacts as the direction of an externally applied magnetic field was varied\cite{sok07}. 

The recent advances in synthesizing low-dimensional nanostructures with curved geometries -- the next-generation nanodevices\cite{ahn06,mei07,par10} -- has
provided us an entirely novel platform of systems whose magnetotransport properties in the ballistic regime are inextricably intertwined with the geometry of the nanostructure. When subject to transversal magnetic fields, the orbital electron motion is governed by the radial field projection, and thus the magnetic field becomes effectively inhomogeneous. Henceforth, the magnetotransport properties of these systems are comparable to those encountered in conventional planar structures in an inhomogeneous magnetic field\cite{mul92,ibr95,ye95,zwe99}. The fact that transport measurements performed on segments of cylinders have shown a notable asymmetry with respect to the external magnetic field direction\cite{fri07,vor07} suggests the possible occurrence of a BAMR effect whose origin does not stem from spin-orbit coupling, but can be rather individuated in the peculiar quantum motion of electrons in curved nanosystems. 

In the current work, we review different incarnations of the geometry-driven BAMR effect in semiconducting low-dimensional systems and its physical motivation.
We start by introducing in Sec.~\ref{sec:quantum} the quantum mechanical properties of electrons constrained to curved manifolds when subject to externally applied magnetic fields. Next, in Sec.~\ref{sec:csn}, we discuss the transport properties of semiconducting core-shell nanowires in transversal magnetic fields with emphasis on the  directional dependence of the low-temperature magnetoresistance. The prediction of a sizable BAMR up to room temperature in rolled-up nanotubes is reviewed in Sec.~\ref{sec:runt}. In Sec.~\ref{sec:strain} we present \textit{new results} on the effect of the inhomogeneous strain distribution, which is inherently present in curved layers, on the transport properties of open tubular structures. Conclusions are drawn in Sec.~\ref{sec:conc}. 

\section{Quantum theory of constrained charged particles subject to external electromagnetic fields}
\label{sec:quantum}
The synthesis of core-shell nanowires\cite{pnt1,pnt2,pnt3,pnt4}, rolled-up nanotubes\cite{runt1,runt2,runt3,runt4,runt5}, and other nanostructures in which quantum carriers are confined to curved geometries, {\it e.g.} carbon nanotubes\cite{cnt1,cnt2,cnt3,cnt4,cnt5}, has renewed the 
 interest in the quantum mechanical properties of particles in curved manifolds. 
The current theoretical paradigm relies on a thin-wall quantization procedure originally introduced by Jensen, Koppe\cite{jk}, and da Costa\cite{costa} (JKC). The quantum motion on a generic two-dimensional (2D) surface is treated as the limiting case of the motion of a quantum particle in an ordinary three-dimensional (3D) space but subject to a confining force acting in the direction normal to the 2D manifold. As a result of the strong confinement, quantum excitation energies in the normal direction are raised far beyond those in the tangential directions, and can be thus safely neglected. It is then possible to deduce an effective dimensionally reduced Schr\"odinger equation in which the effect of the curved space is entirely encoded in a so-called quantum geometric potential\cite{costa}. The thin-wall quantization procedure has been widely employed in recent years\cite{jkc1,jkc2,jkc3,jkc4,jkc5,jkc6,jkc7,jkc8}. 
Furthermore,  the experimental realization of an optical analog of the curvature-induced geometric
potential has provided empirical evidence for the validity of this squeezing procedure\cite{jkc-exp}.

The JKC thin wall-quantization procedure can also be applied in presence of external  electric and magnetic fields\cite{prlsch,ortixsch}. We will now review the analytical derivation of the corresponding Schr\"odinger equation valid for a generic 2D geometry, which thus describes in the most appropriate way the quantum mechanical properties of carriers in real curved semiconducting nanostructures subject to externally applied fields. We start out with the Schr\"odinger equation minimally coupled with the four-component vector potential in a generic curved three-dimensional space. Adopting Einstein summation convention, it reads

\begin{align}
i\hbar\bigg[\partial_t-\frac{iQA_0}{\hbar}\bigg]\psi = -\frac{\hbar^2}{2m^\star}G^{ij}\bigg[{\cal D}_i-\frac{iQA_i}{\hbar}\bigg]\bigg[{\cal D}_j-\frac{iQA_j}{\hbar}\bigg]\psi,
\label{eq:sch3D}
\end{align}
where $Q$ is the particle charge, $m^\star$ is the effective mass of the carriers, $G^{ij}$ is the inverse of the metric
tensor $G_{ij}$, while $A_i$ are the covariant components of the vector potential $A$ with the scalar potential defined by $V = -A_0$.
The covariant derivative ${\cal D}_i$ is defined by ${\cal D}_i v_j = \partial_iv_j - \Gamma^k_{ij}v_k$, where $v_j$ are the covariant components of a 3D vector field ${\bf v}$, 
and $\Gamma^k_{ij}$ is the affine connection related to the 3D metric tensor by 
$$\Gamma^{k}_{ij}=\dfrac{1}{2} G^{kl}\big[\partial_j G_{li}+\partial_i G_{lj}-\partial_l G_{ij}\big].$$

To proceed further, we now define a coordinate system, and consider a 2D manifold ${\cal S}$ with parametric equation ${\bf r}={\bf r}(q_1, q_2)$.  The portion of 3D space in the immediate neighborhood of ${\cal S}$ can be then parametrized as
${\bf R}(q_1, q_2) = {\bf r}(q_1, q_2)+q_3\hat{N}(q_1,q_2)$, where $\hat{N}$ is the unit vector normal to ${\cal S}$.
The relations between the 3D metric tensor $G_{ij}$ and the 2D surface metric tensor $g_{ij}$ can be found as 
\begin{align}
\notag
G_{ij} = g_{ij}+\big[\alpha g+(\alpha g)^T\big]_{ij}q_3+(\alpha g\alpha^T)_{ij}q_3^2~~~~~i, j = 1, 2,\\
\notag
G_{i3} = G_{3i} = 0~~~~i = 1, 2;~~~~~~G_{33}=1,
\end{align}
where $\alpha$ indicates the Weingarten curvature tensor of the surface ${\cal S}$.
We recall that the Weingarten curvature tensor defines the mean curvature $M$ and Gaussian curvature $K$ of the surface ${\cal S}$ via 
$M = {\rm Tr}(\alpha)/2$ and $K = {\rm Det}(\alpha)$. 

Next, we employ the thin-wall quantization procedure of Da Costa\cite{costa}, and take into account the 
effect of a confining potential  $V_\lambda(q_3)$, where $\lambda$ is a squeezing parameter controlling the strength of the potential.
When $\lambda$ is large, the total wavefunction will be localized in a narrow range close
to $q_3 = 0$. This allows to take the $q_3 \rightarrow 0$ limit in the Schr\"odinger equation Eq. (1),
 thereby defining the effective Schr\"odinger equation in the portion of the 3D space close to the surface ${\cal S}$. It is given by: 
\begin{align}
\notag
i\hbar\bigg[\partial_t-\frac{iQA_0}{\hbar}\bigg]\psi =& -\frac{\hbar^2}{2m^\star}g^{ij}\bigg[d_i-\frac{iQA_i}{\hbar}\bigg]\bigg[d_j-\frac{iQA_j}{\hbar}\bigg]\psi\\
&-\frac{\hbar^2}{2m^\star}\bigg[\partial_3-\frac{iQA_3}{\hbar}\bigg]^2\psi-\frac{\hbar^2}{m^\star}M\partial_3\psi+\frac{iQ\hbar}{m^\star}MA_3\psi+V_\lambda\psi,
\label{eq:sch3D2}
\end{align}
where we introduced the 2D covariant derivatives of the surface metric $g_{ij}$  defined as $d_iv_j = \partial_i v_j-\gamma^k_{ij}v_k$. Here $v_k$ and $\gamma^k_{ij}$ respectively indicate the covariant
components of a generic 2D vector field, and the affine connection related to the 2D metric tensor. 

In Eq. (\ref{eq:sch3D2}), the term $M\partial_3\psi$ yields a coupling among the transversal fluctuations of the wave function and the surface curvature,
while the term $QMA_3$ represents an anomalous curvature contribution to the orbital magnetic moment of the carriers\cite{prama}.
Both these terms, however, vanish in the effective Schr\"odinger equation for a well-defined surface wavefunction\cite{ortixsch}. 
The latter can be defined by introducing a new 3D wavefunction $\chi(q_1, q_2, q_3)$  such that in the event of separability, the surface density probability reads $\chi^{S}(q_1, q_2) \times \int d q_3  | \chi^{N} (q_3) |^2$. Conservation of the norm then requires 
\begin{align}
\psi(q_1,q_2,q_3) = \big[1+2 M q_3+ K q_3^2 \big]^{-1/2}\chi^{N}(q_3)\chi^{S}(q_1,q_2).
\label{eq:totalwf}
\end{align}

After inserting the wavefunction above into Eq. (\ref{eq:sch3D2}) and applying a gauge transformation such that $A_3 \equiv 0$, 
one reaches a separability of the dynamics along the direction normal to the surface $\cal{S}$ from the tangential ones. 
With this, the quantum dynamics on the surface is regulated by the following effective dimensionally reduced Schr\"odinger equation: 
\begin{align}
i \hbar \, \partial_t  \chi^S = -\frac{\hbar^2}{2m^\star}g^{ij}\bigg[d_i-\frac{iQA_i}{\hbar}\bigg]\bigg[d_j-\frac{iQA_j}{\hbar}\bigg]\chi^S+QV\chi^S+V_{S}\chi^S,
\label{eq:hs}
\end{align}
where the last term is the curvature-induced purely quantum geometric potential (QGP)  
reading  
\begin{align}
V_{S} = -\frac{\hbar^2}{2m^\star}(M^2-K).
\label{eq:qgp}
\end{align}
The QGP can strongly influence the electronic structure, as well as the electron transport properties of nanosystems with complex curved geometric shapes. 
It has been shown, for instance, that in nanotubes with Archimedean spiral shaped cross sections, the QGP results in the appearance of bound state localized close to the maximum curvature regions, and whose number coincides exactly with the spiral winding number\cite{runt}.

\section{Magnetotransport in Core-Shell Nanowires}
\label{sec:csn}

Core-shell nanowires (CSN) are recently conceived semiconducting heterostructures \cite{pnt1,pnt2,pnt3,pnt4}. They are made of a thin semiconducting layer (shell) surrounding a core in a tubular geometry. This heterostructure can be engineered in such a way that carriers are confined to the shell. This is the case for instance of GaAs/InAs CSN\cite{hpnt} where the conductive InAs shell surrounds the GaAs core. Commonly, CSN have hexagonal cross-sections \cite{hpnt}, but also triangular\cite{tpnt} and circular\cite{cpnt} cross-sections have been achieved. 

The tubular radius of these nanostructures depends on the selection of materials and the growth conditions\cite{pnt4}, but generally lies in the range of a few tens of nanometers. This is much larger than the few nanometer radius of carbon nanotubes \cite{cnt1,cnt2,cnt3,cnt4,cnt5}. On the one hand, this allows to neglect the atomistic details of CSN, thereby making an effective mass description as developed in Sec.~\ref{sec:quantum} reliable. On the other hand, the effect of external applied magnetic fields are much stronger in CSN since the magnetic length of weak/intermediate fields is comparable to the curvature radius. As a result, the quantum states of the carriers are determined by the interplay between these two length scales. In carbon nanotubes a similar regime can be reached only in high magnetic fields\cite{cnt-hall,cnt-abeffect}. 
Below, we discuss the rich variety of quantum states that arise in the tubular shell of CSNs when a transversal magnetic field is applied.

\subsection{Quantum states of a cylindrical electron gas in a transversal magnetic field}
Before applying the effective mass description reviewed in Sec.~\ref{sec:quantum} to the tubular two-dimensional electron gas formed in the shell of CSNs, we first discuss the possible characteristic trajectories an electron can trace in a conducting channel under the action of a generic magnetic field in the ballistic regime, {\it i.e.} when the characteristic dimensions of the channel are smaller than the electronic mean free path. Assuming a constant magnetic field in the direction perpendicular to the channel plane, the classical electron trajectory can be classified as a cyclotron orbit, skipping orbit and transversing orbit depending on whether the trajectory collides with zero, one and both the walls of the channel\cite{skip2deg}. The cyclotron orbits correspond quantum mechanically to states in Landau levels and are realized whenever the cyclotron diameter is smaller than the channel width. 
Transversing orbits occur instead  in the opposite regime and correspond to quantum states in magnetoelectric subbands. Finally, skipping orbits correspond to edge states. There is, however, a fourth electron trajectory that can be encountered in a conducting channel subject to a magnetic field. This can be classified as a snake orbit: it consists of two cyclotron semiorbits of opposite chiralities joined together, and corresponds to the propagation of electrons parallel to the boundary between two magnetic domains. Therefore snake orbits, and the corresponding quantum mechanical snake states, arise only in highly inhomogeneous magnetic fields\cite{mul92,ibr95,ye95,zwe99}. Snake states have drawn considerable attention in recent years since they have been experimentally observed in graphene\cite{snakegra1,snakegra2}. 

It has been recently pointed out\cite{nt-snake1,nt-snake2} that snake states can arise using the geometry of CSNs 
without necessitating for an inhomogeneous magnetic field. They are formed indeed in presence of constant magnetic fields, and thus coexist with Landau states. 
To show this, we consider a CSN with circular cross section. The ensuing cylindrical two-dimensional electron gas (C2DEG) formed in the shell can be parametrized as usual as  $\textbf{r}(\phi,z)= R \cos\phi~\hat{i} +R \sin\phi~\hat{j}+ z~\hat{k}$, where $R$ is the curvature radius of the cylinder while $\left\{\hat{i},\hat{j},\hat{k}\right\}$ is the triad of conventional Cartesian unit basis vectors. 
The covariant components of the surface metric tensor are then given by $g_{\phi\phi} = R^2,~g_{zz} = 1$ and $g_{\phi z} = g_{z\phi} = 0$. With this, Eq.~\ref{eq:hs} yields the following effective two-dimensional Hamiltonian 
\begin{align}
{\cal H}=\frac{1}{2m^{\star}}\Bigg[\Big(\frac{\hbar}{i} \frac{1}{R}\partial_{\phi}- eA _{\phi} \Big)^{2}+\Big(\frac{\hbar}{i}\partial_{z}-eA_{z}\Big)^{2}\Bigg] -\frac{\hbar^{2}}{8m^{\star}R^2}. 
\label{eq:hc2}
\end{align}
The last term in the equation above is the QGP, which corresponds to a homogeneous chemical potential shift since the curvature radius is constant. For a generic homogeneous magnetic field, the vector potential in the gauge required for the separability of the quantum dynamics along the tangential components of the cylindrical surface from the normal one is\cite{prlsch,nt1} 
$\left\{A_\phi, A_z, A_{q_3}\right\} = \left\{B_{\parallel} R /2, B_{\perp}R\sin(\phi-\theta), 0\right\}$,
where $B_{\parallel}$ accounts for the magnetic field component along the tube axis, while $B_{\perp}$ accounts for a generic transversal component, {\it i.e.} 
$B_{\perp}=|B_{\perp}| (\cos\theta\,\hat{i}+\sin\theta\,\hat{j})$. 

\begin{figure}[tbp]
\begin{center}
\includegraphics[width=.8\textwidth]{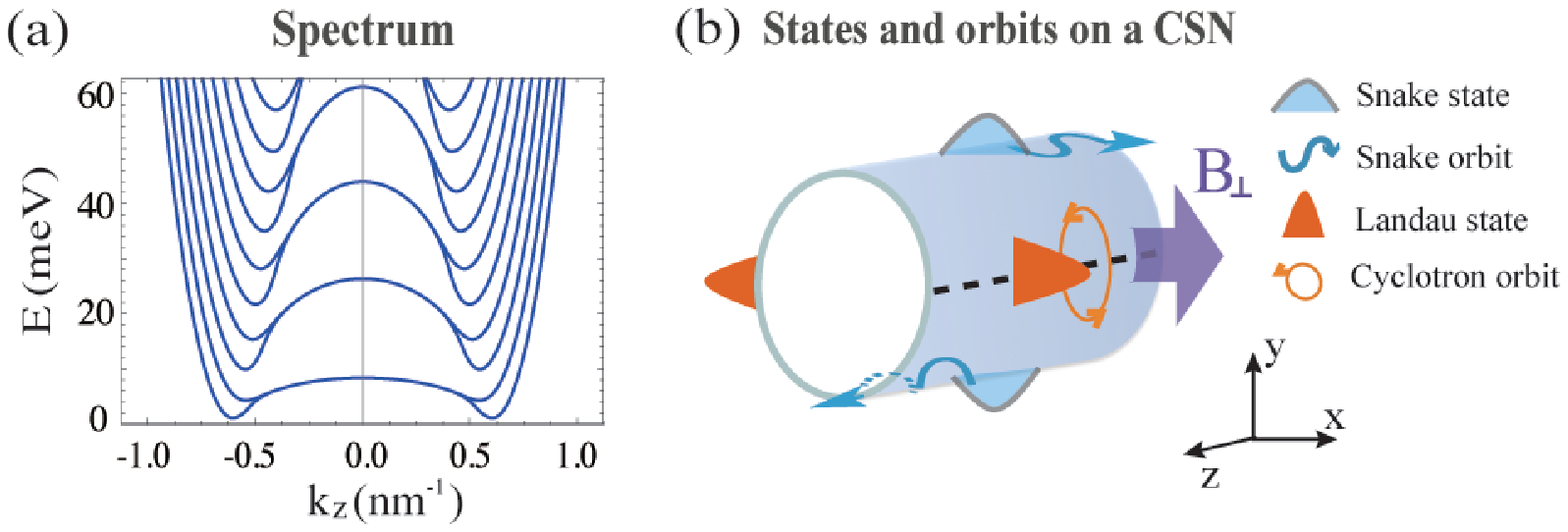}
\caption{(a) Energy spectrum of the 2DEG formed in the shell of a CSN of radius $40$ nm radius subject to a transversal field of 10 T magnitude. (b) Sketch of the corresponding electron trajectories realized at different momenta along the translationally invariant tube axis direction.} 
\label{fig:CNT}
\end{center}
\end{figure}

\begin{figure}[tbp]
\begin{center}
\includegraphics[width=.7\textwidth]{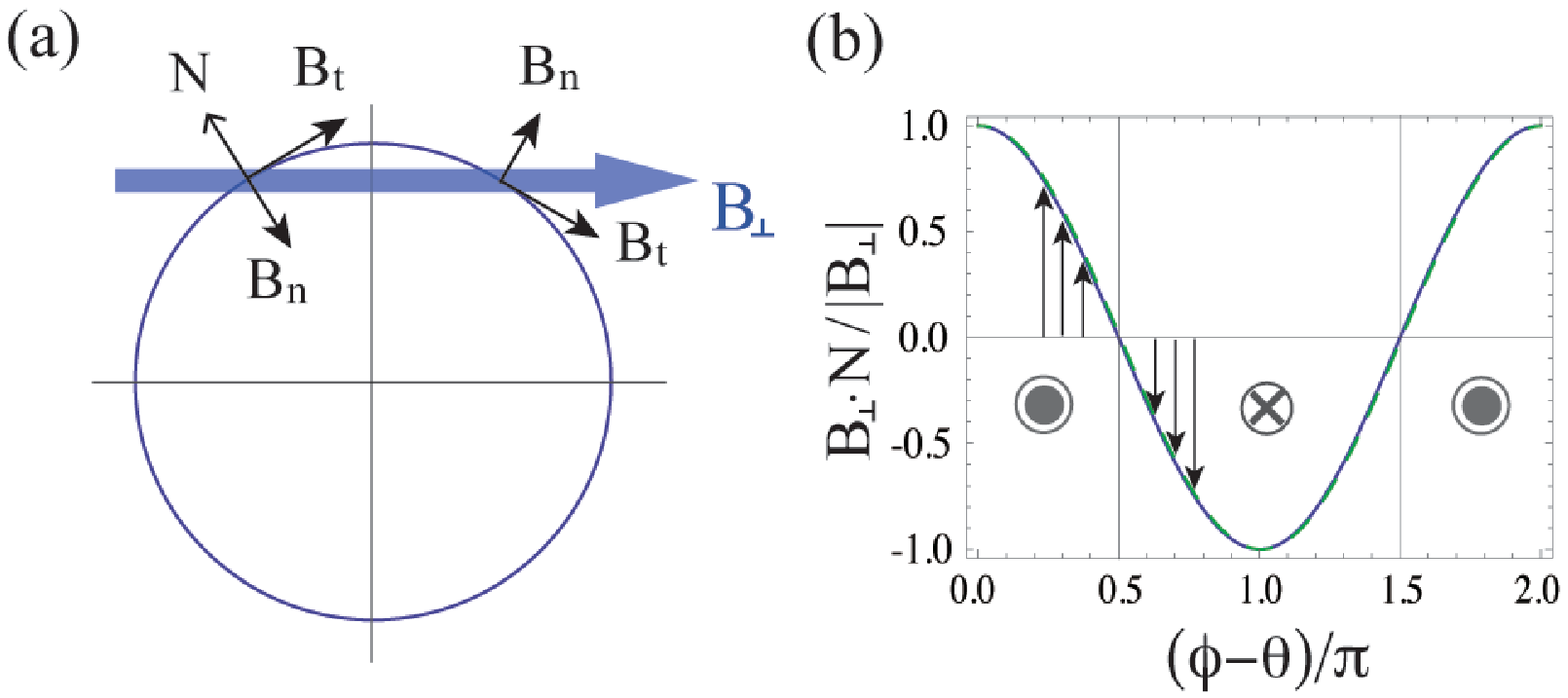}
\caption{(color online) (a) Circular cross section of a CSN  with $\phi \in \left[0, 2 \pi \right] $ subject to a transverse magnetic field $\textbf{B}_{\perp}$. 
$\textbf{B}_{n},\textbf{B}_{t}$ are the components along the surface normal and the surface tangential direction respectively. (b) Behavior of the oscillatory normal component of the magnetic field responsible for the formation of snake orbits.} 
\label{fig:figsupp1}
\end{center}
\end{figure}

Since we are interested in the electronic states appearing under the action of a perpendicular magnetic field, we set the longitudinal component $B_{\|}=0$. Fig.~\ref{fig:CNT}(a) shows the energy spectrum of a C2DEG of $40$ nm radius subject to an external magnetic field of $B_{\perp} \simeq 10$ T. At small values of the momentum $k_z$ along the translationally invariant tube axis direction, the spectrum displays two degenerate ``flat" subbands corresponding to quasi-one-dimensional Landau-like states. The smallness of the Landau magnetic length $l_B =\sqrt{\hbar/(eB)}$ as compared to the C2DEG radius of curvature $R$ indeed allows for the formation of cyclotron orbits centered at the two positions where the tangential plane of the CSN is orthogonal to transversal magnetic field direction [c.f. Fig.~\ref{fig:CNT}(b)]. By increasing the momenum $k_z$, however, the quasi-one-dimensional Landau subbands start to acquire a characteristic parabolic dispersion [c.f. Fig.~\ref{fig:CNT}(a)]. 
We can identify these dispersive states as snake states which are centered, due to Lorentz force, at the two points where the tangential plane of the CSN is parallel to the transversal magnetic field direction [c.f. Fig.~\ref{fig:CNT}(b)]. 
The appearance of snake states follows from the fact that the carrier trajectories are entirely determined by the projection of the transversal magnetic field along the CSN surface normal $B_{\perp}\cos(\phi-\theta)$. In the immediate vicinity of the two points where the transversal magnetic field direction is orthogonal to the surface normal, this ``effective" inhomogeneous magnetic field switches sign, thereby realizing the conditions for the formation of snake orbits [c.f. Fig.~\ref{fig:figsupp1}].

\begin{figure}[tbp]
\begin{center}
\includegraphics[width=.7\textwidth]{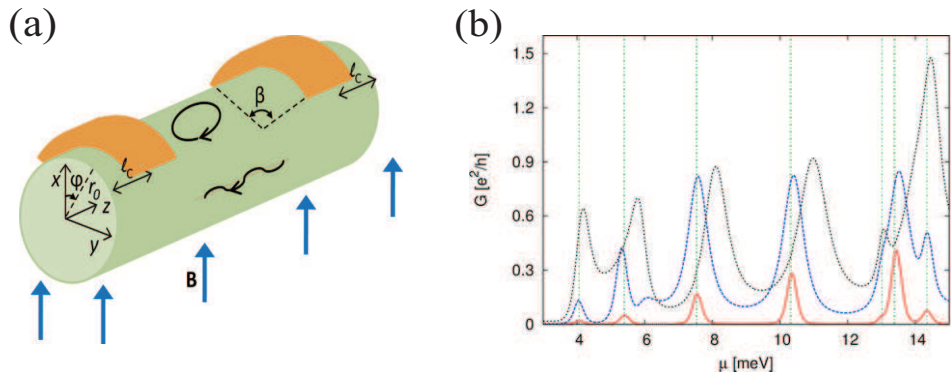}
\caption{(color online) (a) Core-shell nanowire in a transverse magnetic field with two contacts placed  over an angular interval $\beta$, centered at distances $l_c$ from the ends of the nanowire. (b) Conductance of the coupled system as a function of Fermi energy $\mu$. The contacts cover an angular interval $\beta=\pi/2$, positioned over a cyclotron region (solid line) or a snaking region (dashed line), with $l_c = 0$ nm in both cases. The dotted curve is obtained with the same parameters as the dashed curve, but with the contacts shifted by $l_c = 11$ nm. Reprinted from Ref.[58] (copyright 2015 American Chemical Society).}
\label{fig:CNT-BAMR}
\end{center}
\end{figure}

The coexistence of Landau and snake states in core-shell nanowires subject to a transversal magnetic field yields very peculiar features in the quantum transport of core-shell nanowires. As shown in Fig.~\ref{fig:CNT}(a), snake states are situated at the bottom of the energy spectrum, and thus they represent the main actors in the quantum transport of CSNs at low chemical potential. By coupling a CSN  to leads via contacts partially covering the shell of a CSN [c.f. Fig.~\ref{fig:CNT-BAMR}(a)], and employing the Green's function formalism to calculate the phase-coherent conductance in the linear transport regime, it has been recently shown\cite{cnt-bamr} that snake states induce isolated peaks in the quantum conductance [c.f. Fig.~\ref{fig:CNT-BAMR}(b)]. This occurs at chemical potentials corresponding precisely to the snake state energies [c.f. the vertical lines in Fig.~\ref{fig:CNT-BAMR}(b)]. 
Moreover, the 
amplitude of the peaks can be tailored by rotating the direction of the externally applied magnetic field: since the contacts cover a finite angle of the nanowire shell, it follows that the conductance peak amplitudes grow with the overlap between the snake states and the contacts, which is maximum when the magnetic field and the central position of the contacts are orthogonal to each other [c.f.Fig.~\ref{fig:CNT-BAMR}(b)].

\subsection{Prismatic core-shell nanowires}
As mentioned above, the high quality free standing nanowires used as substrates for the epitaxial overgrowth of core-shell structures do not in general possess a circular cross section. This is because the stability of specific crystallographic surfaces leads to faceting. Specifically, III-V based core nanowires often grow as hexagonal crystals. Henceforth, the overgrown shell retains the $n$-fold symmetry of the nanowire used as a substrate.  Transmission electron microscopy images\cite{tem1} show that each edge of the nanostructure actually consists of a region extending over a few nanometers with finite curvature. This, in turn, implies that the cross section of a prismatic CSN can be seen as a polygon where the edges are rounded and possess a constant curvature over a finite length. The attractive QGP then results in a series of square wells with the rounded edges of a prismatic CSN that tend to be regions of preferred localization\cite{gaasnt1}. It has been thereby demonstrated  that when the curvature-induced quantum wells are $\simeq$ 10 meV deep, the ground state of the system in the absence of external fields consists of a set of tunnel-coupled one-dimensional channels  located at the edges of the CSN. This geometry-induced localization is strongly sensitive to the effect of a transversal magnetic field, with a marked directional dependence. A magnetic field orthogonal to one of the facets of the prismatic CSN leads to the formation of low-energy snake states localized at the rounded edges orthogonal to the facet. Therefore the effect of the magnetic field is to enhance localization in two side edges of the CSN while the other edges are charge depleted. On the other hand, for a magnetic field pointing toward one edge of the CSN, the $\mathcal{D}_{6h}$ of the nanostructure implies that the formation of the low-energy snake states occurs along the facets of the CSN thereby resulting in a charge depletion of all edges. Similar features do occur also for the higher energy Landau states localized at the bottom and top region of the CSN with respect to the field direction. If these regions correspond precisely to two edges of the CSN, localization is enhanced at those two edges, while the other four edges are charge depleted. On the other hand, if the bottom and top region of the CSN correspond to two facets of the CSN, all the six edges are charge depleted. 

\begin{figure}[tbp]
\begin{center}
\includegraphics[width=.7\textwidth]{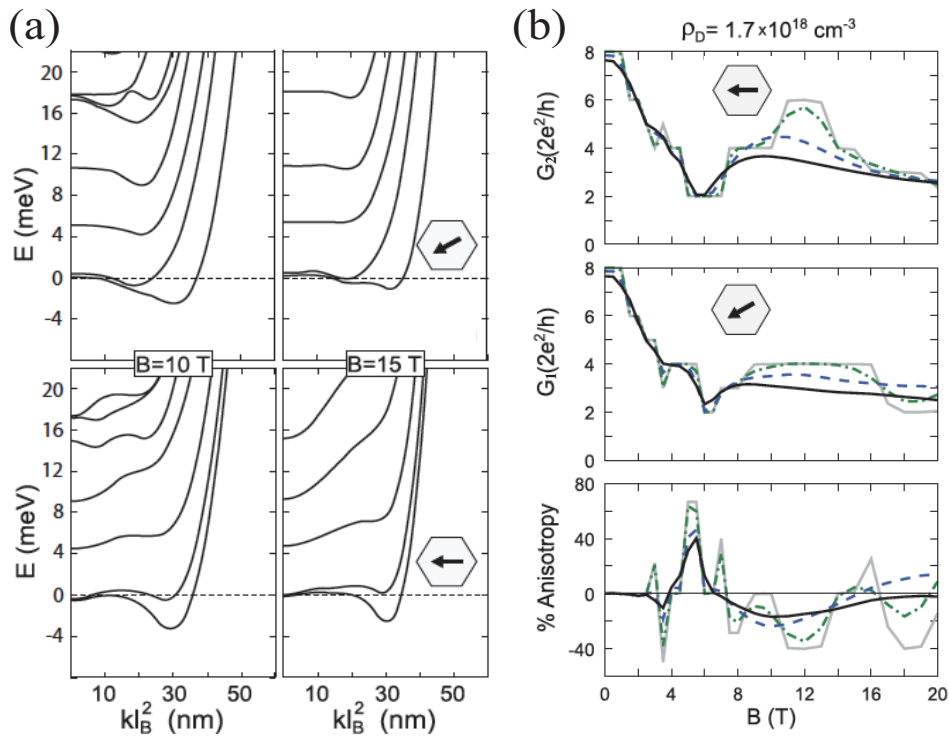}
\caption{(color online) (a) Magnetic spectra of a prismatic core-shell nanowires with a radius of 40 nm and electron density $\rho \simeq 1.7 \times 10^{18} cm^{-3}$. 
The horizontal dashed lines show the Fermi
energy position. (b) (Top and Middle) Magnetoconductances obtained in two representative fields and at selected temperatures: 
0.01 (solid gray lines), 0.5 (dashed-dotted lines), 2 (dashed lines), and 4 K (solid lines). 
(Bottom) Magnetoanisotropy calculated by $2(G_1-G_2)/(G_1+G_2)$.
Reprinted from Ref.[61] (copyright 2013 American Physical Society).} 
\label{fig:PNT-BAMR}
\end{center}
\end{figure}

Such an inhomogeneous localization of the electron gas has been later confirmed by fully three-dimensional self-consistent field calculations of realistic prismatic CSN subject to a transversal magnetic field, and including the effect of electron-electron interactions\cite{pnt-bamr}. In particular, it has been shown that in the high-density regime, the dominating electron-electron interaction leads to a localization at the prismatic core-shell interface which can be largely affected by transversal magnetic fields $B_{\perp} \simeq 10$ T. This gives rise to a directional dependent complex band dispersion [c.f. Fig.~\ref{fig:PNT-BAMR}(a)], eventually leading to an anisotropy of the magnetoconductance with respect to the field orientation [c.f. Fig.~\ref{fig:PNT-BAMR}(b)]. This sizeable ballistic anisotropic magnetoresistance rapidly decreases with temperature since the directional dependence of the band dispersions lies in the meV regime. As a result, a robust BAMR effect in prismatic CSN can only occur in the sub-Kelvin temperature range.

\section{Rolled-up Nanotubes}
\label{sec:runt}

The rolled-up nanotechnology -- thin solid films that curl up after being partially released from a substrate\cite{runt1,runt2,runt3,runt4,runt5,review-arch1,review-arch2} -- allows to manufacture open tube nanostructures, hereafter named rolled-up nanotubes (RUNTs). 
RUNTs offer exciting possibilities due to the exceptional design flexibility. Controlling the thickness of the thin film layers allows to achieve tubular diameters ranging from several microns down to few nanometers\cite{review-exp,review-exp2,review-exp3}, but also the number of windings can be precisely controlled. 
Furthermore, tubular structures belonging to different material classes can be manufactured, ranging from semiconductors\cite{si} to semiconductor/oxide bilayers \cite{si-sio}, metallic/semiconducting multilayers\cite{metal}, and even semiconducting/organic multilayers\cite{organic}. As a result, new nanodevices based on RUNTs have been fabricated. These include, but are not limited to,  field effect transistors\cite{fet},  magnetic sensors\cite{sensor}, energy-storage elements\cite{battery}, and optical microcavities \cite{omn,omp}.

\subsection{Magnetic states on a rolled-up nanotube}

\begin{figure}[tbp]
\begin{center}
\includegraphics[width=.9\textwidth]{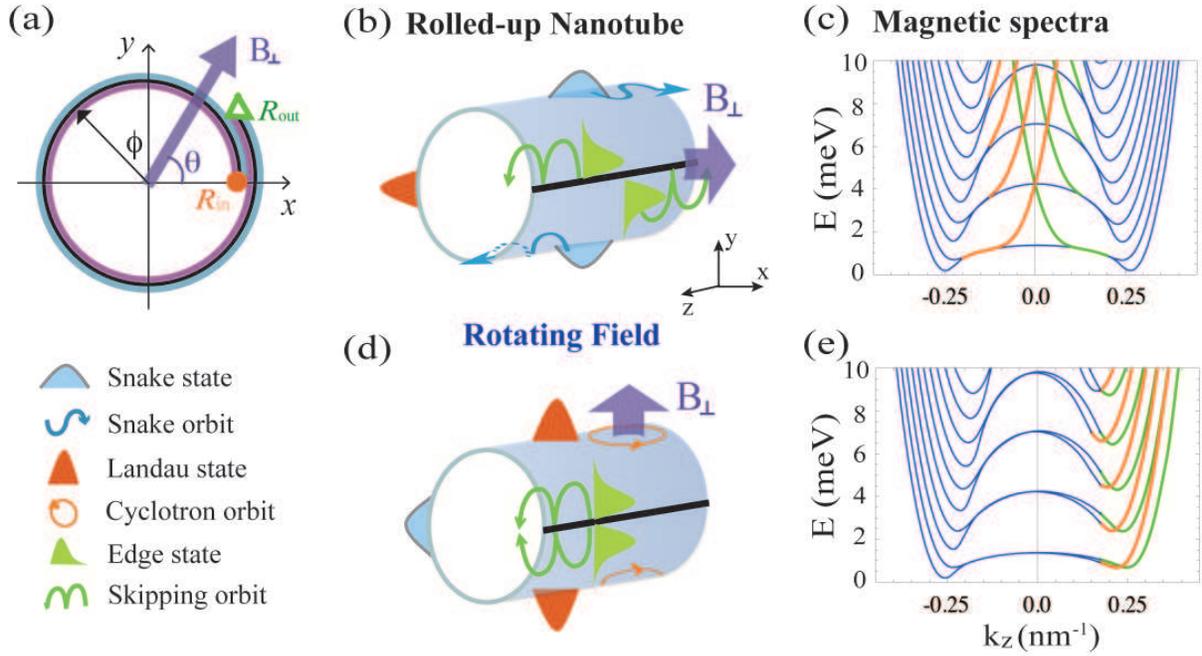}
\caption{(color online) (a)  Cross section of an Archimedean spiral shaped RUNTs with winding number $w=1.1$.  
(b,d) Characteristic electron trajectories formed in a RUNT with an externally applied magnetic field pointing toward the inner open edge (b) and orthogonal to it (d). (c,e) show the corresponding magnetic spectra with the coexistence of quasi-one-dimensional Landau states, snake states and edge states. The inner radius of the RUNT is $R_{in} \equiv 32 \pi$ nm $\simeq 100$ nm. The transversal magnetic field strength has been set to $B_{\perp}=1.65$T. }
\label{fig-runt}
\end{center}
\end{figure}

Although single material nanostructures have been proposed\cite{singlematerial}, RUNTs are generally made out from bilayer or multilayer thin films of different materials. For GaAs-based heterojunctions, this implies the existence of a two-dimensional electron gas in a tubular geometry whose cross section can be fairly approximated by an Archimedean spiral [c.f. Fig.~\ref{fig-runt}(a)]. The corresponding parametric equation of the nanostructure in cylindrical coordinates then reads $\{r,\phi,z\}=\{l\phi,\phi,z\}$, where $l$ controls the distance between successive windings of the spiral, and mixed periodic and open boundary conditions have to be imposed, {\it i.e.} $z \in (-\infty, \infty)$ while $\phi \in (\phi_{in}, \phi_{out})$\cite{runt}. The total number of windings of the RUNTs is determined by $w=(\phi_{out}-\phi_{in})/(2\pi)$. 

The effective dimensionally reduced Schr\"odinger equation introduced in Sec.~\ref{sec:quantum} then yields the following Hamiltonian for the curved two-dimensional electron gas in the presence of an externally applied magnetic field: 
\begin{equation}
{\cal H}=\frac{1}{2m^{\star}}\Bigg[\Big(\frac{\hbar}{i} \frac{1}{\sqrt{g_{\phi\phi}}}\partial_{\phi} -  e A_{\phi} \Big)^{2}+\Big(\frac{\hbar}{i}\partial_{z}-eA_{z}\Big)^{2}\Bigg]-\frac{\hbar^{2}}{8m^{\star}}\alpha^{2}_{\phi\phi},
\label{eq:hs2}
\end{equation}
where we introduced the metric tensor and Weingarten curvature tensor components  
$g_{\phi\phi}=l^{2}(1+\phi^{2})$ and $\alpha_{\phi\phi}=(2+\phi^{2})/[l(1+\phi^{2})^{3/2}]$.
The covariant components of the vector potentials corresponding to an homogeneous magnetic field are\cite{prlsch} 
$\left\{A_{\phi}, A_z, A_{q_3}\right\} = \left\{l B_{\parallel}\sqrt{1+\phi^{2}}/2, B_{\perp}l\phi\sin(\phi-\theta), 0 \right\}$
where, as before, $B_{\parallel}$ indicates the magnetic field component along the tube axis while $B_{\perp}$ 
is the transversal component of the magnetic field. 
In the  $\phi_{in} \gg 1$ regime, which is satisfied whenever the inner radius of the nanotube $R_{in}= l \phi_{in}$ is much larger than the separation between windings, the covariant components of the vector potential can be approximated as the ones of a C2DEG. Henceforth, one expects the magnetic states in a RUNT with $R_{in} \gg l$ to be equal in nature to the states realized in a core-shell nanowire of circular cross section, but with open ends along the azimuthal direction [c.f. Fig.~\ref{fig-runt}(b)]. 

To show this, we first notice that in the effective Hamiltonian, the longitudinal component of the magnetic field can be gauged away using the transformation on the wavefunction $\Psi(\phi,z) \rightarrow {\rm Exp}[\frac{i}{\hbar}e\int d\phi A_{\phi}]\Psi(\phi,z)$. In addition, since the transversal component of the magnetic field does not break the translational symmetry along the tube axis, the wave function can be separated as $\Psi(\phi,z)=\chi(\phi)\times e^{ik_{z}z}$, where $k_{z}$ is the momentum the $z$ direction. The ensuing one-dimensional Hamiltonian for $\chi(\phi)$ then reads: 
\begin{align}
{\cal H}_{1D}=-\frac{\hbar^{2}}{2m^{\star}}\frac{\partial_{\phi}}{\sqrt{g_{\phi,\phi}}}\bigg(\frac{\partial_{\phi}}{\sqrt{g_{\phi,\phi}}}\bigg)
-\frac{\hbar^{2}}{4}\alpha^{2}_{\phi\phi}+{\cal V}_{P},
\label{eq:h1d}
\end{align}
with ${\cal V}_{P}$ corresponding to the magnetic potential
\begin{equation}
{\cal V}_{P}=\frac{1}{2m^{\star}}\left[\hbar k_{z}-eB_{\perp}l\phi\sin\big(\phi-\theta\big)\right]^{2}.
\label{eq:vp}
\end{equation}

Fig.~\ref{fig-runt}(c) shows the magnetic spectrum for a GaAs-based (effective mass $m^{\star} = 0.067 m_e$) RUNT with inner radius $R_{in} \approx 100$ nm and a winding number $w=1.1$ [c.f. Fig.~\ref{fig-runt}(a)]. In addition, the strength of the externally applied magnetic field has been set to $B_{\perp}=1.65$T, while its direction corresponds to $\theta=0$ in Fig.~\ref{fig-runt}(a), {\it i.e.} the magnetic field points toward the inner open end of the tubular structure. In perfect analogy with core-shell nanowires with circular cross section, snake states for finite values of momentum $k_z$ are formed at the bottom of the energy spectrum. However, for $k_z \simeq 0$ one can identify a single non-degenerate quasi-one-dimensional Landau-like states coexisting with two dispersive edge states which are localized at the inner and outer radius of the RUNT. These dispersive edge states are a unique feature of the open tubular geometry of the nanostructure: one of the two cyclotron orbits formed in the closed geometry of CSN fractionalizes, due to the hard walls, into two skipping orbits which acquire a clear dispersion along $k_z$ [c.f. Fig.~\ref{fig-runt}(b)]. 

Since the open curved geometry of a spiral RUNTs breaks the rotational symmetry of the embedding three-dimensional space, one also expects a strong interplay between the location of the hard wall boundaries and the direction of the externally applied transversal magnetic field. This is explicitly manifested in Fig.~\ref{fig-runt}(e) where we show the magnetic spectrum considering the magnetic field direction $\theta=\pi/2$, {\it i.e.} the magnetic field direction is perpendicular to the inner edge axis. The magnetic spectrum strongly resembles the spectrum of a CSN with circular cross section [c.f. Fig.~\ref{fig:CNT}] with the following exception: the snake states at the bottom of the energy spectrum with $k_z > 0$ are substituted by dispersive edge states. This, however, only leads to an asymmetry in the energy spectrum due to the different nature of the magnetic states at $k_z<0$ (snake states) and $k_z>0$ (edge states). 

The features of the magnetic spectra  just reported persist up to a critical energy $E_c$ given by the characteristic Landau level energy splitting $\hbar \omega$, with $\omega$ the cyclotron frequency, renormalized by a geometrically tuneable factor $R_{in}^2 / (2 \, l_B^2)$. The existence of this critical energy can be obtained using a purely classical analysis based on electron trajectories. The existence of quasi-one-dimensional Landau levels is preserved indeed as long as the radius of the cyclotron orbit $k_F l_B^2$ -- with $k_F$ the Fermi momentum  --  is smaller than the curvature radius. For $k_F l_B^2 > R_{in}$ indeed, transversing orbits colliding with the two open edges of the tubular structure appear\cite{skip2deg}. Therefore, independent of the magnetic field direction, one observes the coexistence of snake and edge states with magnetoelectric subbands. For the set of parameters discussed above, the threshold between the two regimes occurs at a characteristic energy $\sim 40$ meV, which is larger than $k_B T$ at room temperature.

\subsection{Ballistic anisotropic magnetoresistance}

The knowledge of the magnetic spectra allows to make interesting predictions on the transport properties of these novel low-dimensional nanostructures. In particular, by neglecting intersubband scattering processes\cite{qpc-book,buttprl,ibm-qpc} the two terminal conductance in the ballistic regime can be derived using the Landauer formula\cite{japan-qc,bag-lg}
\begin{equation}
G(E_F, T,\theta)= \int_{0}^{\infty} G(E,0, \theta) \,\dfrac{\partial f}{\partial E_F} \, d E, 
\end{equation}
where $f$ indicates the Fermi-Dirac distribution, $E_F$ is the Fermi energy while $G(E,0,\theta) = 2 e^2 N(\theta) / h $ is the conductance at zero temperature proportional to the number $N (\theta)$ of occupied magnetic subbands for a given magnetic field direction $\theta$. 

\begin{figure}[tbp]
\includegraphics[width=.7\columnwidth]{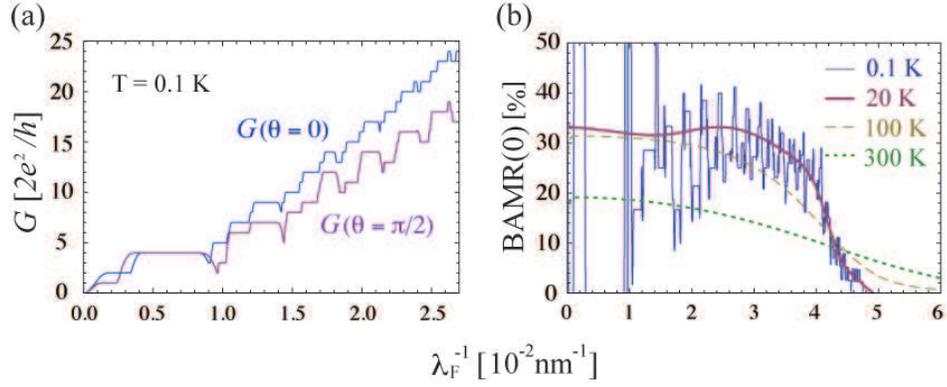}
\caption{(Color online). (a) Low-temperature magnetoconductance measured in $2 e^2 / h$ units as a function of the inverse of the 2DEG Fermi wavelength for a magnetic field oriented parallel to the edge axis (blue line) and perpendicular to it (purple line). (b) Maximum value of the anisotropy BAMR$(0)$ for temperatures up to room temperature. Reprinted from Ref.[82].}
\label{fig:fig3}
\end{figure}

Fig.~\ref{fig:fig3}(a) shows the behavior of the low-temperature magnetoconductance as a function of the inverse of the Fermi wavelength of the 2DEG $\lambda_F=\sqrt{2 \pi / n}$, with $n$ indicating, as usual, the electron density of the 2DEG. Owing to the effective one-dimensionality of the nanostructure, the magnetoconductance shows the characteristic step-like increase. Moreover, 
the different features of the magnetic spectra reported in Fig.~\ref{fig-runt} render a substantial anisotropic magnetoresistance\cite{runtbamr} whose magnitude is defined as 
$$\text{BAMR}(\theta)=\dfrac{G (\theta) - G(\pi/2)}{G(\pi/2)}$$
measured from the reference direction perpendicular to the edge axis where the ballistic conductance takes its minimum value. 
Fig.~\ref{fig:fig3}(b) demonstrates the relevance of the BAMR effect showing that its maximum magnitude at $\theta=0$ extends over a wide range of $\lambda_F$, and persists down to the critical Fermi wavelength $\lambda_F^{c} \simeq 2 \pi  l_B^2 / R$ ($ \simeq 25 \,$ nm for the set of parameters introduced above) where the cyclotron orbit radius balances the radius of curvature. 

\begin{figure}[tbp]
\begin{center}
\includegraphics[width=.7\columnwidth]{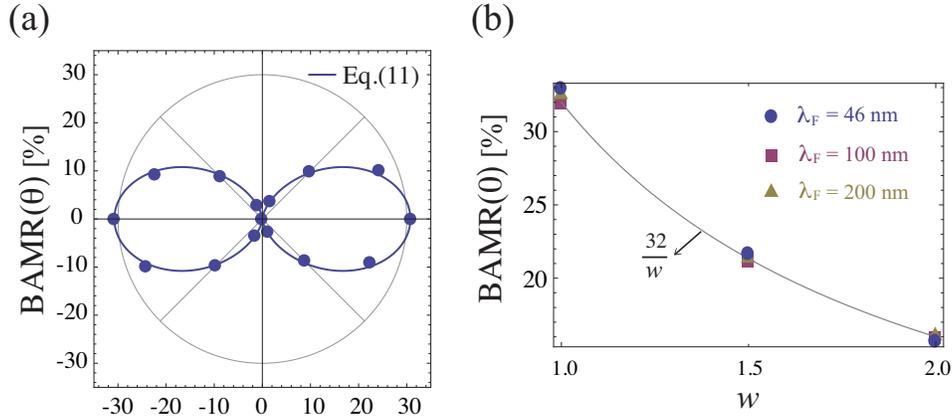}
\caption{(color online). (a) Angular dependence of the BAMR at $T=20$ K  and Fermi wavelength $\lambda_F \simeq 46$ nm. The solid lines corresponds to a fit obtained using  Eq.~\ref{eq:eqang}. (b) Scaling of the BAMR effect as a function of the number of windings of the nanotube for different values of the Fermi wavelength. Reprinted from Ref.[82].}
\label{fig:fig4}
\end{center}
\end{figure}

The characteristic steps in the conductance disappear at temperatures $T > \hbar \omega / (4 k_B)$, {\it i.e.} when the width of the thermal smearing is larger than the subband splitting at the Fermi level. However, this does not immediately imply the concomitant absence of magnetoresistance anisotropy. The rise of the BAMR effect is indeed directly linked to the presence of quasi-1D Landau states characterized by cyclotron orbits, one of which, by tilting the magnetic field from $\theta=\pi/2$ to $\theta=0$ fractionalizes into two conducting edge states. The presence of Landau states occurs below the critical energy $E_{c}$, which is much larger than the subband splitting $\simeq 3$ meV. Since $E_c$ is comparable to $k_B T$ at at 300 K, it follows that a sizable BAMR survives at room temperature as shown in Fig.~\ref{fig:fig3}(b). Moreover, by increasing temperature, the angular dependence of the BAMR [see Fig.~\ref{fig:fig4}(a)] can be accurately described by assuming the functional form of the resistivity for the classical anisotropic magnetoresistance effect\cite{mcg75} 
\begin{equation}
\rho(\theta)=\rho(0) + \left[\rho(\pi/2) - \rho (0) \right] \sin^2{\theta}.
\label{eq:eqang}
\end{equation} 

It must be also emphasized that the BAMR effect of non-magnetic semiconducting rolled-up nanotube is a generic robust physical phenomenon, which is independent of geometric details. By extending the analysis presented above to the case of RUNTs with winding number $w=1.5,2$, it has been shown\cite{runtbamr} that the anisotropy in the magnetoconductance scales as $1/w$, as can be simply understood considering that the ratio between the number of edge states and the number of snake Landau states is inversely proportional to the winding number. 

\section{Strain effects on the electronic and magnetotransport properties of Rolled-Up NanoTubes}
\label{sec:strain}

The rolling of  semiconducting layers is a mechanical self-assembly process, which is due to the relaxation of elastic stresses. However, it is worth noticing that the final rolled-up structure is not strain-free. Because of the curved layers, the strain has a distribution profile along its depth\cite{review-arch1,review-arch2}. Such a nanoscale variation of strain leads to considerable band-edge shifts with regions under tensile and compressive strain shifting in opposite direction\cite{runt-strain}.
The coupling of inhomogeneous strain distributions to the electronic structure of low-dimensional curved nanostructures has been 
put on a formal basis recently\cite{strain}. As we will review below, nanoscale variations of strain render a strain-induced quantum geometric potential, which is of the same functional form of the QGP but strongly, often gigantically, boosting it. In this section, we will show how this strain-induced geometric potential affects both the electronic and the magnetotransport properties of semiconducting RUNTs. 

\subsection{Strain-induced  geometric potential }

For a generic bent nanostructure with tubular geometry, the stress-free surface ${\cal S}$ of the curved layer can be parametrized as ${\bf r} = {\bf r}(s,z)$,
where $z$ is the coordinate along the tubular axis,
while $s$ is the arclength along the curved direction of the surface ${\cal S}$ measured from an arbitrary reference point. The three-dimensional portion of space of the curved layer can be then parametrized as ${\bf R}(s,z,q_3) = {\bf r}(s,z) + q_3 \hat{N}(s)$  with $\hat{N}(s)$ the unit vector normal to ${\cal S}$. The strain distribution can be evaluated assuming the plane strain condition, which states that the strain component along the tubular axis can be neglected when the latter represents the largest structural dimension. With this, it follows that the strain component in the curved direction of the surface varies linearly across the thin film as $\epsilon_{s}=-q_3 \kappa(s)$, where $\kappa(s)$ is the principal curvature. The linear deformation potential theory\cite{deformV1,deformV2} then guarantees the existence of a strain-induced shift of the conduction band corresponding to a local potential for the conducting electrons $V_\epsilon = \gamma q_3 \kappa(s)$ with $\gamma > 0$, 
which yields an attraction toward regions under tensile strain. 
The characteristic energy scale $\gamma$  is proportional to the shear and
the hydrostatic deformation potentials and is expected to lie in the eV scale for conventional semiconductors\cite{deformV1,deformV2}. In the absence of externally applied fields, the time-independent Schr\"odinger equation in the 3D portion of space of the curved layer then reads 
\begin{equation}
-\frac{\hbar^2}{2m^\star}G^{ij} {\cal D}_i {\cal D}_j \psi + V_\epsilon(s,q_3) \psi = E \psi. 
\label{eq:sch3D-strain}
\end{equation}
The presence of the strain-induced local potential in the equation above does not allow for a rigorous separability between the quantum dynamics in the tangential and normal directions. However, the thin-wall quantization procedure can be still employed 
assuming a method of adiabatic separation among fast and slow quantum degrees of freedom\cite{strain}, as sketched below. 

Within the adiabatic approximation, the Schr\"odinger equation Eq.~\ref{eq:sch3D-strain} can be solved assuming the ansatz for the wavefunction 
\begin{equation}
\psi(s,z,q_3) = \big[1- \kappa(s) q_3\big]^{1/2}\chi^{N}(s,q_3) \times \chi^{S}(s,z), 
\label{eq:totalwf-strain}
\end{equation}
where the normal wavefunction $\chi_N$ solves at fixed arclength $s$ the effective one-dimensional Schr\"odinger equation for the fast normal quantum degrees of freedom 
\begin{equation}
E_i^N(s)\chi^N = -\frac{\hbar^2}{2m^\star}\partial_3^2\chi^N+V_\lambda(q_3) \chi_i^N+\tilde{\gamma}(s)\kappa(s) q_3 \chi_i^N. 
\label{eq:hn-strain}
\end{equation}
Here $i$ indicates the transversal subband index and  $\tilde{\gamma}(s) = \gamma - \hbar^2[ \kappa^2(s)+\partial_s\kappa(s)/\kappa(s) ]/(4m^\star)$ is the energy scale of the deformation potential locally renormalized by curvature effects. It is worth pointing out that for typical nanostructures with radii of curvature in the hundreds of nanometer scale, the local curvature-induced renormalization of $\gamma$ can be safely neglected, {\it i.e.} $\tilde{\gamma}(s) \simeq \gamma$. 

The one-dimensional Schr\"odinger equation Eq.~\ref{eq:hn-strain} supplemented by Dirichlet boundary conditions can be solved straightforwardly using second-order perturbation theory, which determines the transversal subband energies $E_i^N(s) = E_i^{0 \, N}+ 2m^\star \delta^4 \gamma^2/\hbar^2 \times f_i \,\kappa(s)^2$, where $\delta$ indicates the thickness of the curved layer, $E_i^{0 \, N}$ are the normal quantum well levels, and the $f_i$'s are numerical constants. 
By subsequently integrating out the $q_3$ quantum degrees of freedom, the effective Hamiltonian for the electron motion along the tubular nanostructures in the first transversal subband reads 
\begin{equation}
{\cal H}^{T}= -\frac{\hbar^2}{2m^\star}\big(\partial^2_s+\partial^2_z\big)-\frac{\hbar^2 \kappa(s)^2}{8m^\star}v_R,
\label{eq:final-strain}
\end{equation}
where the last term corresponds to the QGP with a renormalized strength $v_R=1+4|f_1|(2m^\star \delta^2\gamma/\hbar^2)^2$. 
The nanoscale variation of strain due to curvature then introduces a strain-induced geometric potential (SGP), which is of the same functional form of the quantum geometric potential, but strongly boosting it \cite{strain}. Moreover, the enhancement  is material dependent since it grows quadratically with the effective mass of the carriers $m^{\star}$. As a result, the SGP is expected to be sizable for instance in silicon nanostructures\cite{si,si-sio} due to their larger effective mass\cite{sim}.

\subsection{Strain-induced bound states formation}
As mentioned previously, the QGP in tubular nanostructure with an Archimedean spiral cross section leads to the occurrence of bound states localized close to the inner radius of the nanotube. The enhancement of the geometric potential due to strain leads in a very natural way to a proliferation of such localized states. 

In the absence of externally applied fields, the band structure for a 2DEG in a RUNT consists of parabolic subbands\cite{runt} $E_n(k_z) = E_n^0 + \hbar^2 k_z^2 / (2 m^{\star})$ where the subband index $n$ and $E_n^{0}$ label the eigemodes and the corresponding eigenergies of the effective one-dimensional Hamiltonian Eq.(\ref{eq:h1d}), which in the limit $\phi_{in} \gg 1$ and taking into account  the SGP reads 
\begin{equation}
\tilde{{\cal H}}^{0}=-\frac{\hbar^{2}}{2m^{\star}}\partial^{2}_{s}-{\textit v}_{R}\frac{\hbar^{2}}{16m^{\star}l}\frac{1}{s}.
\label{eq:hamhydrogen}
\end{equation}
Here, $s$ is the arclength of the Archimedean spiral measured from $\phi=0$ 
\begin{equation}
s(\phi)=\frac{l}{2}\Big[\phi\sqrt{1+\phi^{2}}+{\rm log}\big(\phi+\sqrt{1+\phi^{2}}\big)\Big].
\label{eq:s}
\end{equation}
Eq.~\ref{eq:hamhydrogen} corresponds to the effective Hamiltonian of a one-dimensional hydrogen atom in the half-space $s\ge 0$ \cite{runt} with a quantum charge $e_{q}=\hbar\sqrt{{\textit v}_{R}}/(4\sqrt{m^{\star}l})$ that is renormalized by the SGP. 
By including Dirichlet boundary conditions at the initial and final arclength values $s_{in,out}= s(\phi_{in,out})$, one finds that the corresponding spectrum is composed of both atomic-like localized bound states and extended standing waves. Moreover, the number of bound states can be determined using the following analysis. 

The Hamiltonian Eq.~\ref{eq:hamhydrogen} admits a zero-energy eigenstate, which, apart from a normalization constant, takes the following form: 
\begin{equation*}
\psi_{0}(s)=\sqrt{\frac{s}{l'}}J_{1}\Big(\sqrt{\frac{s}{2l'}}\Big)+C\sqrt{\frac{s}{l'}}Y_{1}\Big(\sqrt{\frac{s}{2l'}}\Big),
\end{equation*}
where $l'=l/{\textit v}_{R}$ is the distance between the spiral windings renormalized by the strain-dependent factor $v_R$,  $J$ and $Y$ indicate the  Bessel functions of the first and the second kind respectively, whereas $C$ is a constant that can be fixed by requiring the wavefunction to meet the Dirichlet boundary condition at the inner radius of the nanotube. Since in the $\phi\gg 1$ regime the arclength has the asymptotic form  $s\sim l\,\phi^{2}/2$ [c.f. Eq.~\ref{eq:s}], the zero-energy wavefunction can be recast in the following form
\begin{align}
\psi_{0}(\phi)\sim\sqrt{\phi'}\Bigg[\cos\bigg(\frac{\phi'}{2}-\frac{3\pi}{4}\bigg)+C\sin\bigg(\frac{\phi'}{2}-\frac{3\pi}{4}\bigg)\Bigg],
\label{eq:simis}
\end{align}
where $\phi'=\sqrt{{\textit v}_{R}}\,\phi$. With this, it follows that the zero-energy wavefunction will meet the Dirichlet boundary condition at the outer radius of the nanotube for $\phi'_{out}=\phi'_{in}+2\pi k$ where $k$ is a positive integer. It then follows \cite{runt} that the number of bound states with negative energy is given by 
\begin{align}
n_{max} \le \sqrt{v_R}\frac{\phi_{out}-\phi_{in}}{2\pi}= \sqrt{{\textit v}_{R}} \times N_{R},
\label{eq:nmax}
\end{align}
where $N_R$ indicates the number of rotations. In the absence of strain effects, the number of curvature-induced bound states corresponds precisely to the number of windings of the spiral cross section of the nanotube. However, this number increase by an order of magnitude considering $v_R \sim 10^2$. 

\subsection{Magnetic states}
Having established the influence of the SGP on the electronic structure in the absence of externally applied fields, we now proceed to analyze the competition between the appearance of strain-induced bound states localized at the inner radius of the nanotube, and the existence of snake, Landau and edge states due to the orbital effect of an externally applied magnetic field. 

\begin{figure}[tbp]
\begin{center}
\includegraphics[width=.7\textwidth]{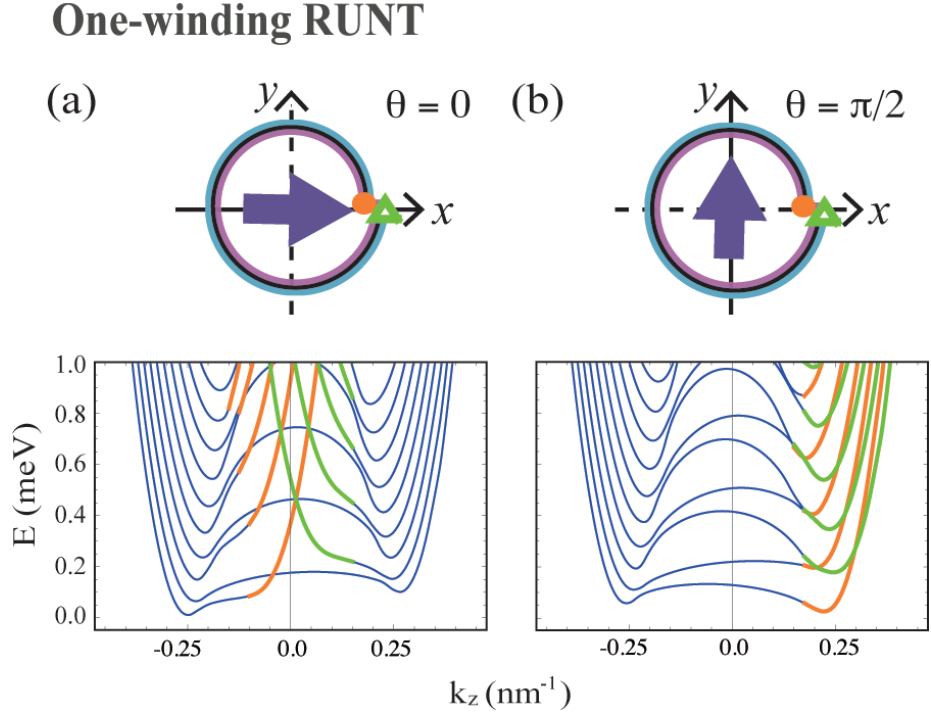}
\caption{(color online) Magnetic spectrum of a single winding rolled-up nanotube subject to an externally applied magnetic field of strength 1.65 T and orientations (a) $\theta = 0$ and (b) $\pi/2$. The strength of the SGP has been set to  $v_{R} = 600$. The inner radius of the nanotube $R_{in}= 32 \pi \simeq 100$ nm.}
\label{fig-sirunt}
\end{center}
\end{figure}

Fig.~\ref{fig-sirunt} shows the magnetic spectrum of a single winding rolled-up nanotube with inner radius $R_{in} \simeq 100$ nm subject to an externally applied magnetic field $B_{\perp}=1.65$ T pointing toward the inner edge of the nanotube [c.f. Fig.~\ref{fig-sirunt}(a)] and orthogonal to it [c.f. Fig.~\ref{fig-sirunt}(b)]. The strength of the SGP has been fixed to $v_R = 600$. 
As compared to the magnetic spectrum obtained in the absence of the SGP [c.f. Fig.~\ref{fig-runt}], it is evident that  strain effects do not change the nature of the magnetic states in this regime. The SGP indeed yields a sizeable energetic separation between  magnetic states located at opposite flanks of the tube. The magnetic spectrum of Fig.~\ref{fig-sirunt}(a) shows an evident asymmetry in the momentum $k_z$ as compared to its partner of Fig.~\ref{fig-runt}(c). The same holds true for Fig.~\ref{fig-sirunt}(b) where the quasi-degeneracy of the Landau states localized at the top and down edges of the RUNT is lifted, and the asymmetry between snake and edge states [c.f. Fig.~\ref{fig-runt}(e)] is enhanced. 

A qualitative change in the magnetic spectrum is encountered when the SGP becomes even more pervasive. This can be accomplished by either decreasing the strength of the externally applied magnetic field or tuning the typical radius of curvature of the tubular structure. In the following, we consider a different RUNT characterized by the same surface area as before, {\it i.e.} the total arclength in the curved direction is kept constant, but with a doubled winding number. For such a nanotube, the values of $\phi_{in,out}$ and hence the radii of curvature reduce dramatically, which eventually leads to an increase in the SGP by an additional order of magnitude. 

\begin{figure}[tbp]
\begin{center}
\includegraphics[width=.7\textwidth]{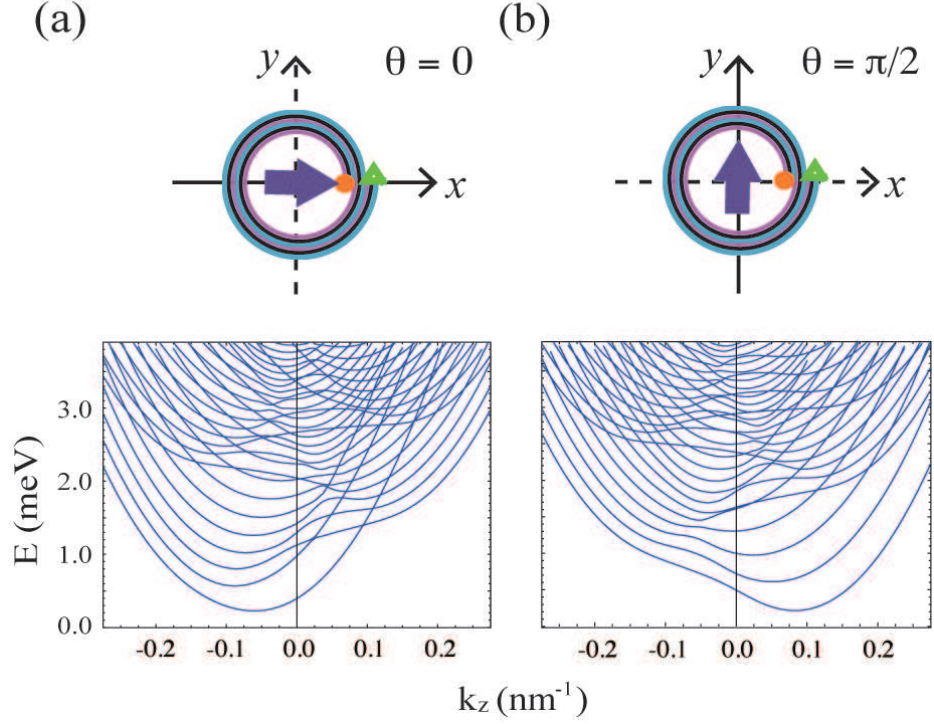}
\caption{(color online) Magnetic spectrum of a two winding RUNT subject to an externally applied magnetic field $B_{\perp}=1.65$ T, and orientations (a) $\theta = 0$ and (b) $\pi/2$. The total surface area of the tube and the strength of the SGP are the same as Fig.~\ref{fig-sirunt}.}
\label{fig-si2runt}
\end{center}
\end{figure}

Fig.~\ref{fig-si2runt} shows the ensuing magnetic spectra where the strength of the external magnetic field has been set, as before, to 1.65 T. Because of the reduced radius of curvature, the formation of cyclotron orbits is precluded and all the magnetic subbands acquire a clear dispersion. Moreover, the low energy part of the magnetic spectrum is all made of states localized in the first turn of the spiral due to the large energy gain caused by the SGP. 

\begin{figure}[tbp]
\begin{center}
\includegraphics[width=.7\textwidth]{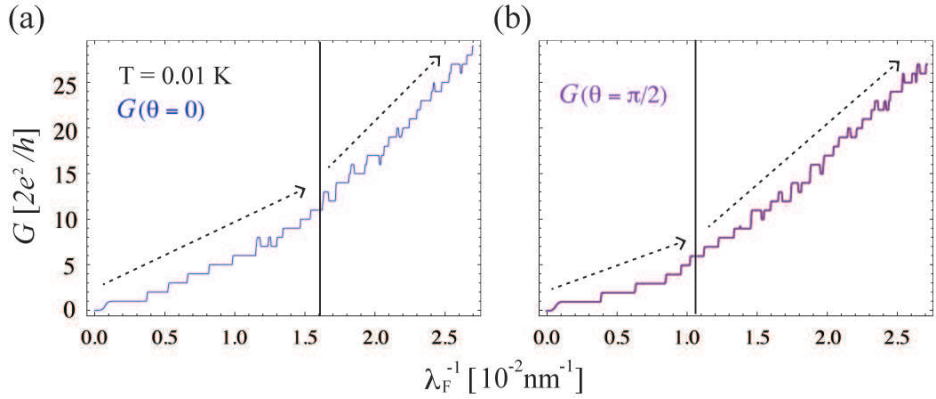}
\caption{(color online) Low-temperature magnetoconductance of a RUNT with $w=2$ as a function of the inverse of the 2DEG Fermi wavelength for a magnetic field oriented (a) parallel to the edge axis and (b) perpendicular to it. The parameter set is the same as Fig.~\ref{fig-si2runt}.}
\label{fig-si2runtg}
\end{center}
\end{figure}

Comparison of Fig.~\ref{fig-si2runt}(a) and Fig.~\ref{fig-si2runt}(b) makes it evident that the magnetic spectrum retains a strong directional dependence. This comes about the energetics of the snake states and it is thus different in origin from the directional dependence in the absence of the SGP. When the magnetic field is oriented in the direction perpendicular to the edge axis, the snake states gain energy due to their larger attractive SGP as compared to the case in which the magnetic field is parallel to the edge axis. Henceforth, lowest energy snake states localized in the second turn of the spiral are occupied at a lower Fermi energy for $\theta=\pi/2$ compared to $\theta=0$. This feature also affects the low-temperature magnetoconductance [c.f. Fig.~\ref{fig-si2runtg}]. Specifically, the step-like increase of the magnetoconductance has a characteristic slope that changes at a critical Fermi wavelength corresponding to the occupation of the snake states localized in the second turn of the spiral [c.f. the vertical lines in Fig.~\ref{fig-si2runtg}]. This critical Fermi wavelength is directional dependent precisely because of the different energetics of the snake states. 

\begin{figure}[tbp]
\begin{center}
\includegraphics[width=.7\textwidth]{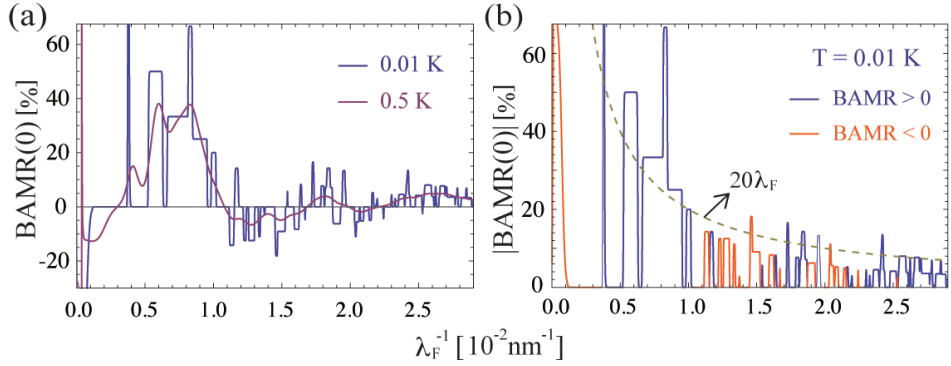}
\caption{(color online) (a) Maximum anisotropy BAMR(0) for a two-winding RUNT in the low-temperature regime. The parameter set is the same as Fig.~\ref{fig-si2runt},~\ref{fig-si2runtg}. (b) Corresponding magnitude of the anisotropy, which decreases with the inverse of the Fermi wavelength.}
\label{fig-si2runtbamr}
\end{center}
\end{figure}

Fig.~\ref{fig-si2runtbamr} shows the corresponding anisotropy in the magnetoconductance. It shows an oscillatory profile which decreases linearly with the Fermi wavelength. As compared to Fig.~\ref{fig:fig3} a sizable anisotropy is found only in the low-temperature regime -- it becomes vanishing small already for a few Kelvin. This follows from the fact that in the present case the anisotropy is not related to the formation of cyclotron orbits but only follows from the directional dependent energetics of the snake states. 

\section{Conclusions}
\label{sec:conc}
The transport properties of semiconductor nanostructures in the presence of external magnetic fields remains a topic of intense research to this day thanks to the enormous experimental advances in manufacturing novel low-dimensional curved systems. Because of the effective inhomogeneous fields felt by the conducting electrons, fascinating physical phenomena arise. In this work, we have reviewed theoretical predictions for the occurrence of directional dependent magnetotransport properties in semiconducting tubular structures. This physical property is widely used in sensor technologies for magnetic recording, electronic compasses, traffic detection, linear and angle sensing, \textit{etc.}, and, to date, was believed to necessitate the concomitant presence of spin-orbit coupling and magnetism even in ballistic nanosystems. The theoretical results reviewed in this work therefore constitute a proof-of-principle demonstration that nanostructures with curved shapes can potentially serve as miniaturized electronic devices with novel functionalities.

\section*{Acknowledgements}
We are grateful to Jeroen van den Brink for a close interaction on the subject of this work. We acknowledge the
financial support of the Future and Emerging Technologies (FET) programme within
the Seventh Framework Programme for Research of the European Commission 
under FET-Open grant number: 618083 (CNTQC).  C.O. has been also
supported by the Deutsche Forschungsgemeinschaft under Grant No. OR 404/1-1.

\end{document}